\documentclass[a4paper,10pt]{amsart}

\usepackage{graphicx,amsaddr}
\usepackage[hang,flushmargin]{footmisc} 
\usepackage{textcomp}
\usepackage{lmodern}
\usepackage{amsmath,amsthm,amssymb}
\usepackage[T1]{fontenc}
\usepackage[english]{babel}
\usepackage[utf8]{inputenc}
\RequirePackage[authoryear]{natbib}
\usepackage{hyperref}

\title{The topology of the directed clique complex as a network invariant}

\author{Paolo Masulli and Alessandro E. P. Villa}

\address{Neuroheuristic Research Group, Faculty of Business and Economics (HEC),
University of Lausanne, Quartier Dorigny, 1015 Lausanne, Switzerland}
\email{paolo.masulli@unil.ch, alessandro.villa@unil.ch}

\begin{document}

\begin{abstract}
We introduce new algebro-topological invariants of directed networks, based on the topological construction of the directed clique complex. The shape of the underlying directed graph is encoded in a way that can be studied mathematically to obtain network invariants such as the Euler characteristic and the Betti numbers. 

Two different cases illustrate the application of the Euler characteristic. We investigate how the evolution of a Boolean recurrent artificial neural network is influenced by its topology in a dynamics involving pruning and strengthening of the connections, and to show that the topological features of the directed clique complex influence the dynamical evolution of the network. The second application considers the directed clique complex in a broader framework, to define an invariant of directed networks, the network degree invariant, which is constructed by computing the topological invariant on a sequence of sub-networks filtered by the minimum in- or out-degree of the nodes.

The application of the Euler characteristic presented here can be extended to any directed network and provides a new method for the assessment of specific functional features associated with the network topology.
\end{abstract}

\renewcommand{\thefootnote}{\fnsymbol{footnote}} 
\footnotetext{\textbf{Keywords:} simplicial homology,
network invariant, recurrent neural network, synfire chain, synaptic plasticity. \newline
\textbf{MSC class:} 05C10, 55-04.}     
\renewcommand{\thefootnote}{\arabic{footnote}}

\maketitle

\section*{Background}

The main interest of algebraic topology is to study and understand the functional properties of spatial structures.
Algebro-topological constructions have been applied successfully in the field of data science 
	\citep{carlsson:topologydata} 
with the application of the framework of persistent homology, which has proved to be a powerful tool to understand the inner structure of a data set by representing it as a sequence of topological spaces. 
A network is a set of points satisfying precise properties of connectedness, which can be used to define a class of topological spaces.
Network theory aims to understand and describe the shape and the structure of networks, and the application of the tools developed within the framework of algebraic topology can provide new insights of network properties in several research fields.

The directed clique complex is a rigorous way to encode the topological features of a network in the mathematical framework of a simplicial complex, allowing the construction of a class of invariants which have only been recently applied for the first time in the context of network theory 
	\citep{giusti:cliquetopology, hess:snowbird}. 
Active nodes are those nodes whose state depend on a set of precise rules that depend on network topology and dynamics.
In a highly interconnected network of such nodes, the activity of each node is necessarily related to the combined activity of the afferent nodes transmitted by the connecting edges. 
Due to the presence of reciprocal connections between certain nodes, re-entrant activity occurs within such network. 
Hence, selected pathways through the network may emerge because of dynamical processes that may produce activity-dependent connection pruning. 
The overall goal of these studies is to understand the properties of a network given the topology described by its link structure.

Neuronal networks are a complex system characterized by coupled nonlinear dynamics. This topic is a long-standing scientific program in mathematics and physics  
	\citep{Guckenheimer90, amit:modeling, Abarbanel:synchronization,Freeman+94jtb}. 
In general, the synchronization of two systems means that their time evolution is periodic, with the same period and, perhaps, the same phase 
	\citep{malagarriga:mesoscopic}.  
This notion of synchronization is not sufficient in a context where the systems are excited by non-periodic signals, representing their complex environment. 
Synchronization of chaotic systems has been discovered 
	\citep{Fujisaka+83,Pikovsky+84,Afraimovich+86} and since then it has become an important research topic in mathematics 
	\citep{Ashwin+94pla}, physics \citep{Ott+94pla} and engineering  
	\citep{Chen99}. 
In interconnected cell assemblies embedded in a recurrent neural network, some ordered sequences of intervals within spike trains of individual neurons, and across spike trains recorded from different neurons, will recur whenever an identical stimulus is presented.  
Such recurring, ordered, and precise interspike interval relationships are referred to as ``preferred firing sequences''. One such example can be represented by brain circuits shaped by developmental and learning processes
	\citep{edelman:topobiology}. 
The application of tools from algebraic topology to the study of these systems and networks will be of great use for determining deterministic chaotic behavior in experimental data and develop biologically relevant neural network models that do not wipe out temporal information 
	\citep{Babloyantz+85pla,Rapp86,Mpitsos+88brb,Celletti+96bc,Celletti+97jsp}

In the current study we introduce a mathematical object, called directed clique complex, encoding the link structure of networks in which the edges (or links) have a given orientation. 
This object is a simplicial complex that can be studied with the techniques of algebraic topology to obtain invariants such as the Euler characteristic and the Betti numbers. 
We propose general constructions valid for any directed network, but we present an application to evolvable Boolean recurrent neural networks with convergent/divergent layered structure 
	\citep{Abeles:corticonics} 
with an embedded dynamics of synaptic plasticity. 
The Euler characteristic, which is defined given the network connectivity, is computed during the network evolution.
We show evidence that this topological invariant predicts how the network is going to evolve under the effect of the pruning dynamics.
Despite being just a toy-example of the dynamics observed in biological neuronal networks, we suggest that algebraic topology can be used to investigate the properties of more refined biologically-inspired models and their temporal patterns. 
We show also that, for a directed network, the Euler characteristic computed on a sequence of networks generated by filtrating its nodes by in- and out-degrees can provide a general metric helpful for a network classification.
Hence, the topological invariants computed for each network in the filtration give a sequence of numbers that may be interpreted as a fingerprint of the complete network.

\subsection*{Acknowledgements}
 
This work was partially supported by the Swiss National Science Foundation grant CR13I1-138032.
We wish to thank Prof. Kathryn Hess, Prof. Ran Levi and Pawe\l{} D\l{}otko for suggesting the idea of using oriented cliques in order to define simplices in a directed graph, as it was shown in the talk given at the SIAM DS15 Conference in Snowbird (USA) in May 2015.

\section*{Results and discussion}

\subsection*{Dynamics of artificial neural networks}

We considered a directed graph representing a simplified model of feedforward neural network with convergent/divergent layered structure with few embedded recurrent connections.
In this model, the nodes represent individual neurons and the connections between them are oriented edges with a weight given by the connection strength.
We have computed the Euler characteristic and its variation during the evolution of such networks, both for the entirety of the nodes in the network and for the sub-network induced by the nodes that are active at each time step in order to detect how the structure changes as the network evolves. 
The Betti numbers and their variation during the network evolution were also computed but we do not discuss further this topological measurement.
Notice that activation of the networks follows a very simple dynamics.
The nodes of the input layer are activated at regular time intervals, which is not meant to be biologically realistic, but has been adopted to favor the simplicity of the model. 
It was shown elsewhere 
	\citep{iglesias:emergencepreferred, iglesias;effectsstimulusdriven} 
that a stable activity level in a network like this could be achieved only with an appropriate balance of excitatory and inhibitory connections.
The networks studied here are oversimplified and formed only by excitatory nodes.
We selected the ranges of the parameters such that the simulations maintained a level of activity for $100$ steps without neither saturation nor extinction of the activity, thus suggesting that connection pruning enabled topological changes.
However, notice that even within selected areas of the parameter space of the simulations we observed that the activity level tended either to increase towards paroxysmal activation (i.e., saturation) or to decrease towards complete inactivation (i.e, extinction).

\begin{figure}
  \centering
  \includegraphics[width=0.8\textwidth]{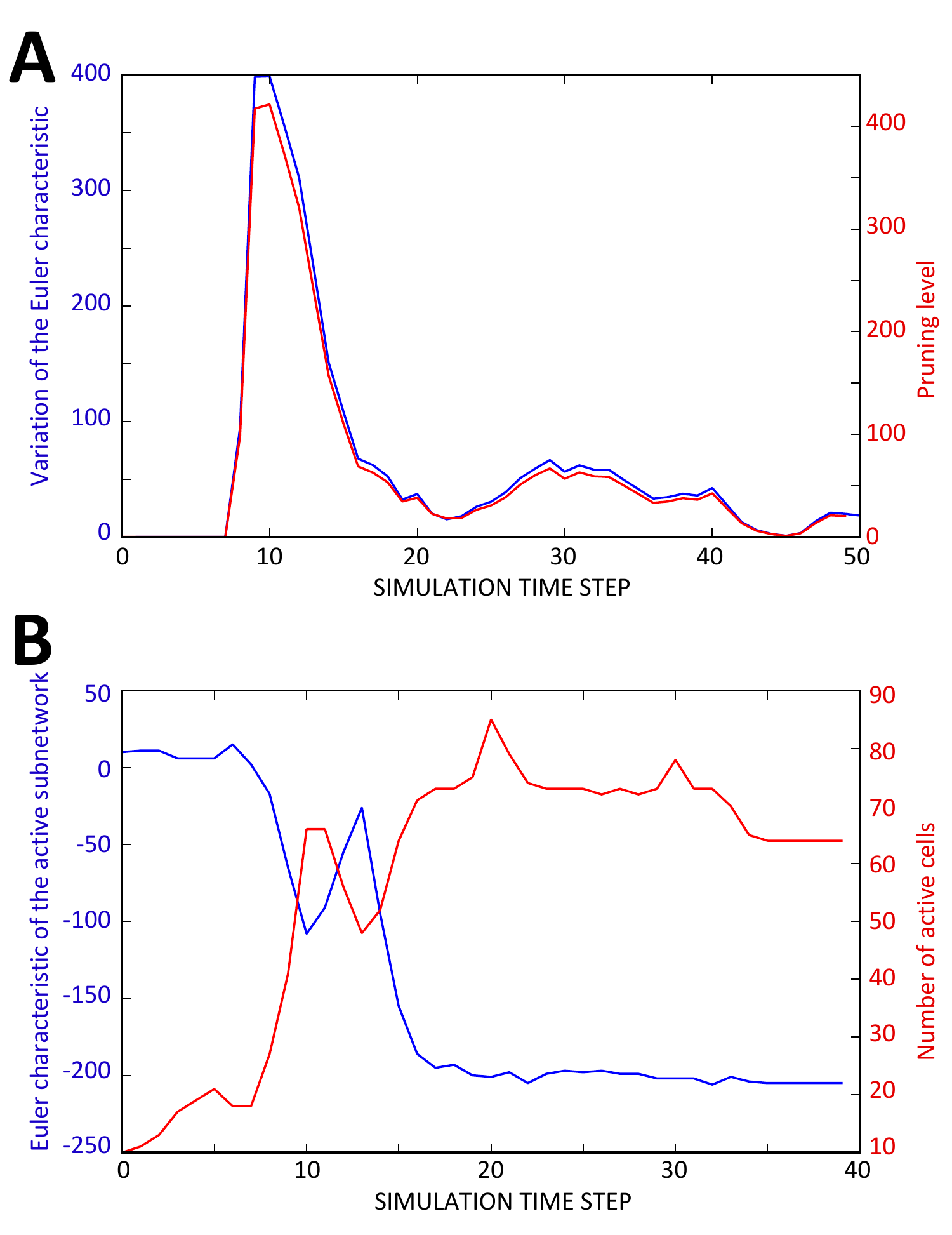}
  \caption{Variation of the Euler characteristic during the neural network evolution.\newline \label{fig:change_euler}
      (A) Differences of the Euler characteristic in subsequent steps of the simulation over time (blue curve) compared to the pruning level (red curve), i.e. the number of pruned connections. 
      Notice that the two curves overlap almost completely, then the variation of the Euler characteristic is a precise measure of the changes in the topology due to the pruning activity.
      \newline
      (B) Evolution of the Euler characteristic of the active sub-network (blue curve) compared to the number of active units during the network evolution (red curve). 
      Notice that the two curves are negatively correlated.
      }
      \end{figure}

We observed that the Euler characteristic of the entire network could detect the pruning activity during the neural network evolution (Figure~\ref{fig:change_euler}A). 
In particular, the step to step variation of the Euler characteristic matched the number of connections pruned over time. 
If we considered only the sub-network of the active nodes, we observed that the Euler characteristic decreased or increased if the number of active nodes increased or decreased, respectively (Figure~\ref{fig:change_euler}B).
Then, the Euler characteristic is a good estimator of the activity level within the network.
These results confirmed that the Euler characteristic gives a precise measure of the topological changes in a network associated with connection pruning for the entire network and associated with activation patterns for the active sub-network.

The type of dynamics undergoing the neural network evolution and the structure of the directed clique complex of that network at the very beginning of the simulation (i.e. before the occurrence of connection pruning) were  correlated.
This was possibly the most unexpected and significant result regarding the dynamics of artificial neural networks.
In the simulations leading to the activation of at least $5\%$ of the nodes, the average number of active units was correlated to the number of simplices, in the directed clique complex, 
of dimension two (Pearson correlation coefficient $r_{(370)} = .560$, $p < 0.001$)  
and dimension three ($r_{(370)} = 0.445$,$p < 0.001$). 
This may appear surprising because the topology of the directed clique complex of a network \textit{a priori} ignores any dynamics of pruning, the evolution of the network topology and how this is going to influence the activation level. 
However, the rationale is that directed cliques are fully connected sub-networks, i.e. sub-networks with an initial and a final node that are connected in the highest possible number of ways.
Then, a high number of directed cliques leads to a higher chance of propagation of the activation through the network. 
Notice the fact that it is essential to consider here only \textit{directed} cliques, because the activation of a node occurs only if the connected upstream nodes are activated.
Activation is indeed a phenomenon happening in a directional way prescribed by the connectivity pattern.
The invariant presented should also be considered as a complementary measurement of complexity for the assessment of the computational power of Boolean recurrent neural networks
	\citep{Cabessa2014e94204}.

\subsection*{Network filtrations and invariants}

The in- and out-degrees of nodes are important factors in shaping the network topology. 
We applied our topological construction to devise invariants for any directed networks. 
We compute the Euler characteristic on a sequence of sub-networks defined by \emph{directed degeneracy} of their nodes, or in other words the in- and out-degrees of the vertices, as described in detail in the Methods section. 
Two separate sequences are defined because in- and out-degrees represent different aspects in the network connectivity.
The sequences of sub-networks of a given network are a \emph{filtration} in the sense that each network appearing in the sequence is contained in all those that follow.
The values of the Euler characteristic for each network of the sequences gives rise to two separate sequences of integers that give a measure of the shape and the topology of the complete network. 
We propose this invariant to describe general directed networks.

\begin{figure} 
  \centering
  \includegraphics[width=0.6\textwidth]{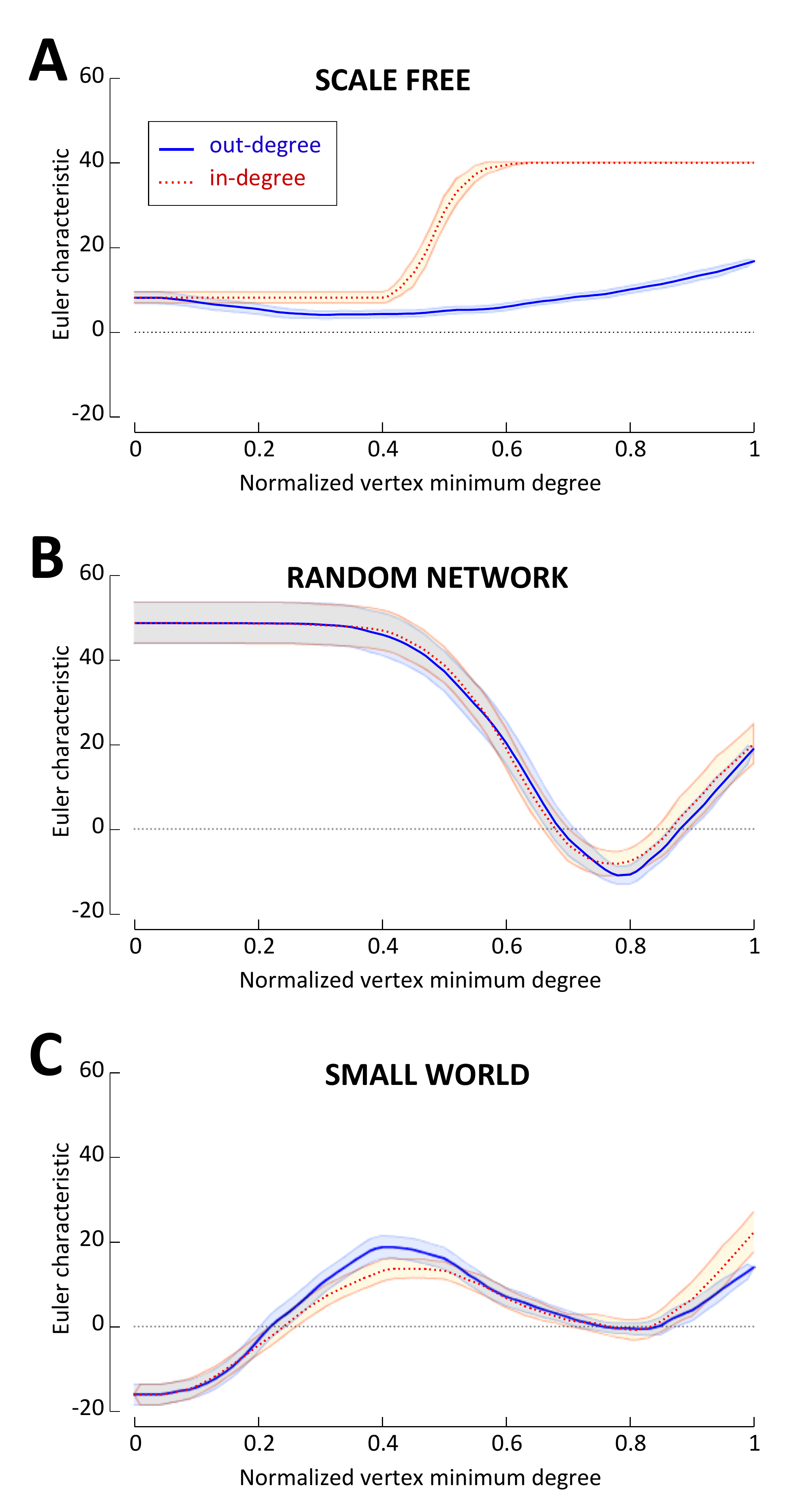}
  \caption{Plots of the degree-filtered Euler characteristic for different networks.\newline \label{fig:filtration_invariant}
  		All networks were generated with $n = 40$ nodes chosen with a fixed ordering.
		(A) Barabasi--Albert Scale Free (SF) network ($m = 10$).
		(B) Erd\H{o}s--R\'{e}nyi Random Network (RN) ($p = 0.2$).
		(C) Newman-Watts-Strogatz Small World (SW) network ($k = 20$, $p = 0.4$).
		\newline The plots show the averaged curves (in-degrees as red dotted lines and out-degrees as blue solid lines) for $N=50$ experiments for each type of network. 
		Confidence bands correspond to $95\%$ intervals. 
		The x-axis is normalized, expressing each filtration level as the degree of the vertices at that level divided by the maximum degree of vertices present in the network. 
This representation makes it possible to compare the plots for different networks. 
Notice that the variation range, the monotonicity and the comparison between the in- and the out-degree filtrations can be used to distinguish the different types of networks.
      }
      \end{figure}

The sequences of the Euler characteristic of the in-degree and the out-degree filtrations are plotted as a function of the normalized minimum degree of vertices
for representative types of a scale-free network (SF) 
	\citep{BarabasiAlbert1999},
a random network (RN)
	\citep{ErdosReny1959,Gilbert1959},
and a small world network (SW)
	\citep{Watts1998,Newman2000}
(Figure~\ref{fig:filtration_invariant}). 
This normalization is necessary to compare networks of different sizes at each filtration level posing the maximum degree of the vertices in the network to 1, as described in the Methods section.
Each network type was simulated 50 times using different random seeds. 
In the case of a SF network the values of the Euler characteristic of the in-degree filtration (dotted red line) is always larger than the curve of the out-degree filtration (solid blue line, Figure~\ref{fig:filtration_invariant}A).
Moreover, for SF networks the curve of the in-degree increases sharply at near $0.4$ and reaches a maximum value of the Euler characteristic approximately at $0.6$ of the normalized maximum value of the vertex minimum degree.
The out-degree curve increases monotonically after this level of vertex minimum degree but does not reach the in-degree curve.
In the case of a RN network the curves of in- and out-degrees overlap at all levels of the filtration (Figure~\ref{fig:filtration_invariant}B). 
It is interesting to notice that the maximum value of the Euler characteristic is observed for the smallest values of the vertex minimum degree.
Then, for RN networks, both curves decrease to a minimum at approximately $0.8$ of the normalized maximum value of the vertex minimum degree, followed by a monotonic increase.
For a SW network both curves of in- and out-degrees start from the minimal value of the Euler characteristic with the least vertex minimum degree, followed by a non monotonic increase and a tendency of overlap between the two curves (Figure~\ref{fig:filtration_invariant}C).
The monotonicity of the curves and the differences between in- and out-degree filtration differ greatly for the three types of networks, thus suggesting that this invariant is a good descriptor of network topology. 

A distinct topological invariant defined for non-directed networks, referred to the Betti curves, was recently proposed by 
	\cite{giusti:cliquetopology} 
following the idea of filtering the network by the weight of connections. 
This invariant appears well suited for continuously distributed connection weights, for instance when the weights are related to the distances of points and represent a symmetric relation between nodes. 
In the case of directed networks with modifiable values of connection weights restricted to a limited set
	\citep{iglesias:dynamics},
the network dynamics evolves towards a bimodal distribution of the connection weights densely grouped near the minimum and maximum values of the range. 
This is a general behaviour in neuronal networks 
	\citep{song:competitive}. 
In this kind of networks, filtering the network by the connection weights following 
	\cite{giusti:cliquetopology} 
is not suitable, because most connections would have the same weight. 
Our approach for directed networks is to filter the connections by the in- and out-degrees separately in order to measure how the nodes of each degree shape the topology of the network.
It is important to point out that other methods are based on spectral properties of the adjacency matrix and therefore only make sense if all the transformations of the network data are linear
	\citep{Brouwer2012bk}. 
%

The results presented here open the way to further applications of the topological invariants.
The analytic study of the values of the Euler characteristic in the filtrations framework can  provide a metric of similarity between networks which is only dependent on their internal topology, thus allowing the application of clustering algorithms for the detection of distinct functional classes of networks.
The study of brain complex networks in clinical neuroscience offers as a particularly promising field of application of the new topological invariant, as suggested by other studies using different techniques to the same aim
	\citep{Stam2007pp92,fallani:2014}.
Another promising application is the study of the temporal dynamics in neural activity. 
The finding of precise and repeating firing sequences in experimental and simulated spike train recordings has been discussed with respect to the existence of synfire chains
	\cite{Abeles1982,Abeles:corticonics}
or chaotic attractors
	\cite{Celletti+96bc,Villa1998pp763}.
In both cases the underlying network structure is assumed to be a directed graph.
This hypothesis together with the assumption of spike-timing modifiable connections provide a rational basis for the application of topological invariants towards understanding the association between topological structures and neural coding.

\section*{Conclusions}

We have developed new invariants for directed networks using techniques derived from algebraic topology, showing that this subject provides a very useful set of tools for understanding networks and their functional and dynamical properties. 
Simple invariants such as the Euler characteristic can already detect the changes in the network topology. 
The promising results shown here are a contribution to the application of algebraic topology to the study of more complex networks and their dynamics, including models of neuronal networks that are biologically inspired.
We believe that the framework present here may open the way to many computational applications to unveil data structures in data and network sciences.

\section*{Methods}

\subsection*{Graphs and clique complexes}

An \emph{abstract oriented simplicial complex} $K$  
	\citep{hatcher2002algebraic} 
is the data of a set $K_0$ of vertices and sets $K_n$ of lists $\sigma = (x_0, \dots, x_n)$ of elements of $K_0$ (called \emph{$n$-simplices}), for $n \geq 1$, with the property that, if $\sigma = (x_0, \dots, x_n)$ belongs to $K_n$, then any sublist $(x_{i_0}, \dots, x_{i_k})$ of $\sigma$ belongs to $K_k$. The sublists of $\sigma$ are called \emph{faces}. 

We consider a finite directed weighted graph $G = (V,E)$ with vertex set $V$ and edge set $E$ with no self-loops and no double edges, and denote with $N$ the cardinality of $V$. Associated to $G$, we can construct its \emph{(directed) clique complex} $K(G)$, which is the directed simplicial complex given by $K(G)_0 = V$ and 
\begin{equation}\label{eq:directed_clique_complex}
K(G)_n = \{(v_0, \dots, v_n) \colon (v_i, v_j) \in E \textrm{ for all } i < j \} \quad \textrm{ for } n \geq1.
\end{equation}
In other words, an $n$-simplex contained in $K(G)_n$ is a directed $(n+1)$-clique or a completely connected directed subgraph with $n+1$ vertices. Notice that an $n$-simplex is though of as an object of dimension $n$ and consists of $n+1$ vertices.

\begin{figure}
  \centering
  \includegraphics[width=0.8\textwidth]{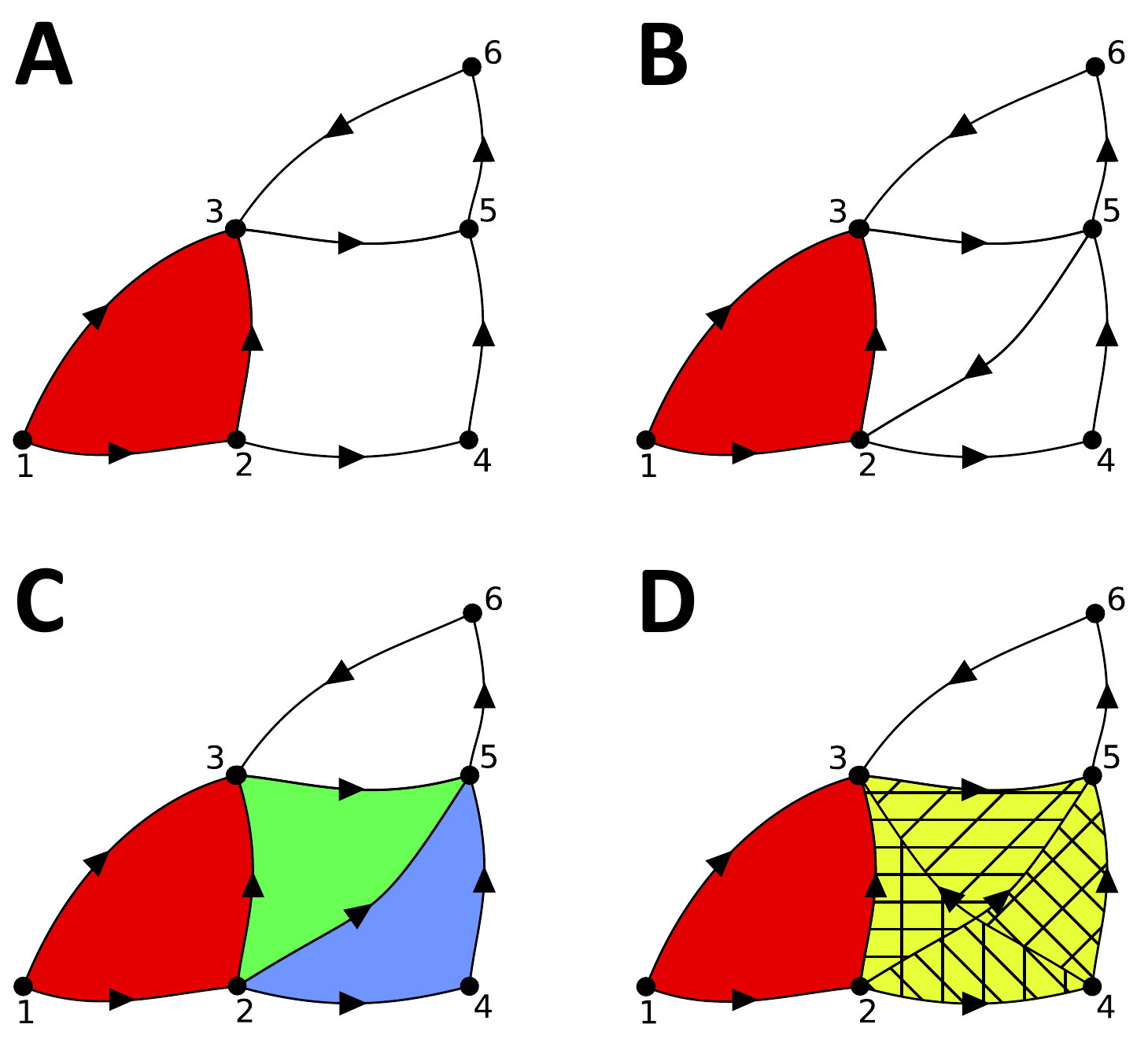}
  \caption{The directed clique complex.\newline  \label{fig:directed_clique_complex}
      (A) The directed clique complex of the represented graph consists of a 0-simplex for each vertex and a 1-simplex for each edge. There is only one 2-simplex (123). Note that '2453' does not form a 3-simplex because it is not \textit{fully} connected. '356' does not form a simplex either, because the edges are not oriented correctly.
      \newline
      (B) The addition of the edge (52) to the graph in (A) does not contribute to creating any new 2-simplex, because of its orientation. The edges connecting the vertices 2, 3 and 5 (respectively 2, 4 and 5) are oriented cyclically, and therefore they do not follow the conditions of the definition of directed clique complex stated in Equation \eqref{eq:directed_clique_complex}.
      \newline
      (C) By reversing the orientation of the new edge (25), we obtain two new 2-simplices: (235) and (245). Note that we do not have any 3-simplex.
      \newline
      (D) We added a new edge (43), thus the sub-graph (2435) becomes fully connected and is oriented correctly to be a 3-simplex in the directed clique complex. In addition this construction gives two other 2-simplices: (243) and (435).
      }
      \end{figure}

By definition, a directed clique (or a simplex in our complex) is a fully-connected directed sub-network (Figure \ref{fig:directed_clique_complex}) this means that the nodes are ordered and there is one source and one sink in the sub-network, and the presence of the directed clique in the network means that the former is connected to the latter in all the possible ways within the sub-network.

\subsection*{The topological invariants}

The directed clique complex is the basic topological object that allows us to introduce invariants of the graph: the \emph{Euler characteristic} of the directed clique complex $K(G)$ of $G$ is the integer defined by
\[
\chi(K(G)) = \sum_{n=0}^N (-1)^n \ \vert K(G)_n \vert,
\]
or in other words the alternating sum of the number of simplices that are present in each dimension.

Let us now consider, for each $n$, the vector space $\mathbf{Z}/2\langle K(G)_n\rangle$ given by the linear combinations of $n$-simplices with coefficients in the field of two elements $\mathbf{Z}/2$. We can define the \emph{boundary maps} $\partial_n \colon \mathbf{Z}/2\langle K(G)_n\rangle \to \mathbf{Z}/2\langle K(G)_{n-1}\rangle$ which are given by mapping each simplex to the sum of its faces. Then we can define the quantities:
\[
\beta_n(K(G)) = \mathrm{dim}(\mathrm{ker}\ \partial_n) - \mathrm{dim}(\mathrm{Im}\ \partial_{n+1}),
\]
given by the difference of the dimension of the space of the $n$-simplices whose boundary is zero and the dimension of the space of boundaries of $(n+1)$-simplices. It can be checked that, if we apply a boundary map twice on any linear combination of simplices, we get zero, and so the quantities $\beta_n(K(G))$ are always non-negative integers. These classically known numbers take the name of \emph{Betti numbers} and, for each $n$, the $n$-th Betti number $\beta_n(K(G))$ corresponds to the dimension of the $n$-th homology space (with $\mathbf{Z}/2$-coefficients) of the clique complex $K(G)$ of $G$. 

The intuitive sense of this construction is to count the ``holes'' that remain in the graph after we have filled all the directed cliques. In particular, the $n$-th Betti number is counting the $n$-dimensional holes. One can also see that $\beta_0$ counts the number of connected components of the graph. A classical result in topology shows a connection between the Euler characteristic and the Betti numbers, expressed by the identity: $\chi(K(G)) = \sum_{n=0}^N (-1)^n \beta_n(K(G))$, which gives another way of computing the Euler characteristic.

Notice that the construction of the directed clique complex of a given network $G$ does not involve any choice, and therefore, since the Betti numbers and the Euler characteristic of a simplicial complex are well-defined quantities for a simplicial complex \citep{hatcher2002algebraic}, our constructions produce quantities that are well-defined for the network $G$, and we shall refer to them simply as the Euler characteristic and the Betti numbers of $G$.

\subsection*{Boolean recurrent artificial neural networks}

\subsubsection*{Network structure and dynamics}

The artificial recurrent neural networks consist of a finite number of Boolean neurons organized in layers with a convergent/divergent connection structure 
	\citep{Abeles:corticonics}.
The networks are composed by $50$ layers, 
each of them with $10$ Boolean neurons.
The first layer is the input layer and all its $10$ neurons
get activated at the same time at a fixed frequency of $0.1$,
i.e. every $10$ time steps of the history. 
Each neuron in a layer is connected to a randomly uniformly distributed 
number of target neurons $f$
belonging to the next downstream layer.
The networks include \emph{recurrence} in their structure, 
meaning that a small fraction $g$
of the neurons appears in two different layers.
This means that a neuron $k$ that is also identified as neuron $l$,
is characterized by the union of the input connections of
neurons $k$ and $l$, as well as by the union of their
respective efferent projections.
%

The state $S_i(t)$ of a neuron $i$ take values $0$ (inactive) or $1$ (active) 
and all Boolean neurons are set inactive at the beginning of the simulation.
The state $S_i(t)$ is a function of the its activation variable $V_i(t)$ 
and a threshold $\theta$, 
such that
$S_i(t)  =  \mathcal{H}(V_i(t)-\theta)$.
$\mathcal{H}$ is the Heaviside function, 
$\mathcal{H}(x)=0 : x<0$, 
$\mathcal{H}(x)=1 : x\ge0$.
At each time step,
the value $V_i(t)$ of the activation variable of the $i^{th}$ neuron
is calculated such that 
$V_i(t+1) =  \sum_{j} S_j(t) w_{ji}(t)$,
where
$w_{ji}(t)$ are the weights of the directed connections 
from any $j^{th}$ neuron projecting to neuron $i$.
The connection weights can only take four values,
i.e.  $w_1 = 0.1$, $w_2 = 0.2$, $w_3 = 0.4$, $w_4 = 0.8$.
At the begin of the simulations all connection weights
are  randomly uniformly distributed among the four possible values.
The weights of all the neurons are computed synchronously at each time step.
%

The network dynamics implements activity-dependent plasticity of the connection weights.
Whenever the activation of a connection does not lead to the activation of its target neuron during an interval lasting $a$ time steps, 
its weight is weakened to the level immediately lower 
than the current one.
Whenever the weight of a connection reaches the lowest level, the connection is removed from the network
	\citep{iglesias:dynamics}. 
Then, the pruning of the connections provokes the selection of the most significant ones and changes the topology of the network.
Similarly, whenever a connection with a weight $w_m$ 
is activated at least $m+1$ consecutive time steps,
the connection weight is strengthened to the level immediately higher 
than the current one.
%
Hence, the parameter space of our simulations was defined by four parameters:
the number $f$ of layer-to-layer downstream connections
in the range $3$--$10$ by steps of 1,
the small fraction $g$ of the neurons appearing in two different layers
in the range $1$--$3$\% by steps of 1\%,
the threshold of activation $\theta$ 
in the range $0.8$--$1.4$ by steps of 0.1,
and the interval $a$ of the weakening dynamics of the connections
in the range $7$--$9$ by steps of 1.

\subsubsection*{Implementation of the simulations}

The simulation software was implemented from scratch in Python.
The network evolved with the dynamics explained above and the program computed the directed clique complex at each change of the network topology.
For the entire network, the directed clique complex was computed each time the connectivity changed because of pruning.
For the sub-network of the active nodes, the computation was carried out at each step of the simulation.

The computed directed clique complexes were used to compute the Euler characteristic 
both for the complexes representing the entire network and for the sub-complexes of the active nodes. 
To compute the directed clique complex of a network we used the implementation of the algorithm of 
	\cite{tsukiyama:cliquealgorithm} 
in the \texttt{igraph} Python package \citep{igraph}, adapted to find directed cliques.
The experiments were run in parallel on several CPUs using the tool GNU Parallel 
	\citep{tange:parallel}.

\subsection*{Network filtrations}

\subsubsection*{Network structures}

Many essential topological features of a network are determined by the distribution of edges over its graph. Different types of distributions result in different types of networks. 
For instance, pure random networks (RN)  
are formed assuming that edges in the network are independent of each other and they are equally likely to occur
	\citep{ErdosReny1959,Gilbert1959}.
For RN we have used the algorithm implemented in the  Python package `NetworkX'
(https://networkx.github.io/)
	\citep{hagberg-2008-exploring}
with the function `erdos\_renyi\_graph' 
with parameters number of nodes $n=40$
and the probability for edge creation $p=0.2$.
These simple construction assumptions are generally not followed in networks
obtained experimentally from ecological or gene systems, telecommunication networks or the Internet which are characterized by short average path lengths and high clustering, resulting in the so called small-world topology (SW)
	\citep{Watts1998,Newman2000}.
For SW we used the same Python package `NetworkX'
	\citep{hagberg-2008-exploring}
with the function `newman\_watts\_strogatz\_graph'
with parameters number of nodes $n=40$
and the number of connected neighbours in ring topology $k=20$
and the probability for adding a new edge $p=0.4$.
Other real-world networks such as brain, social networks, power grids and transportation networks exhibit topologies where more connected nodes, hubs, are more likely to receive new edges.
The presence of these hubs and a power law distribution for the degree of the nodes defines scale-free networks (SF)
	\citep{BarabasiAlbert1999}.
For SF we used the same Python package `NetworkX' 
	\citep{hagberg-2008-exploring}
with the function `barabasi\_albert\_graph'
with parameters number of nodes $n=40$
and the number of edges to attach from a new node $m=10$.

\subsubsection*{Network degree invariant}

Given a directed network $G$, we define two filtrations by sub-networks (ordered sequences of networks in which each network is a sub-network of all the following ones) using the in- and out-degree of nodes. 
Let $ODF(G)$ be the out-degree filtration of $G$: the $i$-th network $ODF(G)_i$ in this filtration is the sub-network of $G$ induced by the vertices having out-degree at least $i$ and all the target nodes of their outgoing connections. 
In the same way we define the in-degree filtration $IDF(G)$: the $i$-th network $IDF(G)_i$ in this filtration is the sub-network of $G$ induced by the vertices having in-degree at least $i$ and all the source nodes of their incoming connections.

We computed the Euler characteristic for each network of the two filtrations, obtaining two sequences of integers, which are plotted to display a measure of the network topology, as a function of the degree levels of the filtration, normalized by the maximum degree present in the network. 
For example, let us consider the case illustrated in Figure \ref{fig:filtration_invariant}B: one of the random networks with $n=40$ vertices 
that we have generated with a parameter $p=0.20$, as described above, had a maximum out-degree of its vertices equal to $19$.
Therefore all the filtration levels have been divided by this value to normalize them (between 0 and 1).

For each network family (SF, RN, SW), we generated $N=50$ distinct networks with different seeds for the random numbers generator (the seeds were uniformly distributed integers in the interval $[1, 10000]$). 
We calculated the network degree filtration invariant sequences for in- and out-degree, which were then averaged for each network family and represented in Figure \ref{fig:filtration_invariant} with the $95\%$ pointwise confidence bands.

\section*{Competing interests}
  The authors declare that they have no competing interests.

\section*{Author's contributions}
   	Conceived and designed the experiments: PM, AEPV.
	Developed the mathematical construction, implemented the simulation, analyzed the results and drafted the manuscript: PM, AEPV.
	Both authors reviewed, read and approved the final manuscript.



\end{document}